\newcommand\DIMPY{(C$_7$H$_{10}$N)$_2$CuBr$_4$}

\documentclass[prl,twocolumn,floatfix,preprintnumbers,amsmath,amssymb,superscriptaddress]{revtex4}

\usepackage{graphicx}
\usepackage{dcolumn}
\usepackage{bm}
\usepackage{color}
\usepackage[usenames,dvipsnames]{xcolor}
\usepackage{ulem}

\begin{document}


\title{Spectrum of a magnetized strong-leg quantum spin ladder}


 \author{D. Schmidiger}
 \affiliation{Neutron Scattering and Magnetism, Laboratory for Solid State Physics, ETH Z\"urich, CH-8093 Z\"urich, Switzerland}
 \author{P. Bouillot}
 \affiliation{{Department of Medical Imaging and Information Sciences, Interventional Neuroradiology Unit, University Hospitals of Geneva, CH-1211 Geneva, Switzerland}}
 \affiliation{Laboratory for Hydraulic Machines, \'Ecole Polytechnique F\'ed\'erale de Lausanne, {CH-1015} Lausanne, Switzerland}
\author{T. Guidi}
\affiliation{ISIS Facility, Rutherford Appleton Laboratory, Chilton, Didcot, Oxon OX11 0QX, United Kingdom}
\author{R. Bewley}
\affiliation{ISIS Facility, Rutherford Appleton Laboratory, Chilton, Didcot, Oxon OX11 0QX, United Kingdom}

  \author{C. Kollath}
 \affiliation{ HISKP, Universit\"at Bonn, Nussallee 14-16,
D-53115 Bonn, Germany}

 \author{T. Giamarchi}
 \affiliation{DPMC-MaNEP, University of Geneva, CH-1211 Geneva,
Switzerland}
 \author{A. Zheludev}
 \email{zhelud@ethz.ch}
 \homepage{http://http://www.neutron.ethz.ch/}
 \affiliation{Neutron Scattering and Magnetism, Laboratory for Solid State Physics, ETH Z\"urich, CH-8093 Z\"urich, Switzerland}
\date{\today}

\ULforem
\begin{abstract}
Inelastic neutron scattering is used to measure the spin excitation spectrum 
of the Heisenberg $S=1/2$ ladder material \DIMPY\ in its entirety, both in 
the gapped spin-liquid and the magnetic field induced Tomonaga-Luttinger 
spin liquid regimes. A fundamental change of the spin dynamics is observed 
between these two regimes. DMRG calculations quantitatively reproduce and help 
understand the observed commensurate and incommensurate excitations. The 
results validate long-standing quantum field theoretical predictions, but 
also test the limits of that approach.
\end{abstract}



\maketitle

One of the main attractions of low-dimensional quantum magnets is that these 
simple systems can be employed as quantum simulators \cite{Feynman1982} for 
fundamental field theories of particle and many body physics 
\cite{Wilczek1998,Ward2013}. Conversely, much of what we know about spin 
chains and ladders is based on continuum quantum field theoretical (QFT)  
non-linear $\sigma$-models (NLSM) \cite{tsvelikbook1,giamarchibook1}. Such 
theories predicted and explained the gapped ``spin liquid'' ground states in 
integer-spin chains \cite{Haldane1983} and even-leg spin ladders \cite{Dagotto1996p1}, 
and provided a unified ``Tomonaga-Luttinger spin liquid'' (TLSL) picture of all 
gapless one-dimensional magnets \cite{Haldane1980}. Moreover, they painted a 
roadmap to how the various phases in real materials are transformed one into 
another through quantum phase transitions \cite{Coldea2010}, and provided new 
insights on the corresponding quantum critical behavior \cite{Lake2010}.

The recent discovery of new materials that experimentally realize spin chain and 
ladder models, as well as progress in neutron spectroscopy techniques and numerical
methods, allow a quantitative verification of the relevant QFT predictions. Moreover, 
one can now push beyond the low-energy limit to explore features of spin chains 
and ladders that are {\it not} covered by continuous mappings of the actual spin 
Hamiltonians. In this context, in the present work, we focus on the quintessential 
model of the strong-leg antiferromagnetic $S=1/2$ Heisenberg spin ladder 
\cite{Barnes1993p1}. We study the spin excitation spectrum on either side of a 
field-induced quantum phase transition between gapped and gapless phases, 
{\it in its entire dynamic range}. An excellent agreement with QFT predictions 
is obtained for some low-energy excitations. At the same time, we uncover 
numerous prominent higher-energy spectral features that were not covered by 
existing field-theoretical approaches.

The basic mechanism of the field-induced transition in a quantum spin ladder is 
well understood through a continuum mapping of the Heisenberg Hamiltonian based 
on Abelian bosonization \cite{Chitra1997,Furusaki1999p1,Konik2002}. It is a 
soft-mode transition, driven by a Zeeman splitting of the lowest-energy magnon 
triplet in the gapped phase. These magnons initially have a ``relativistic'' 
dispersion relation with a mass $\Delta$, and can be viewed as $S=1$ confined 
states \cite{Shelton1996p1} of elementary $S=1/2$ excitations called {\it spinons} 
\cite{Faddeev1981}. Beyond the critical field $H_{\mathrm{c}1}=\Delta/g\mu_\mathrm{B}$, 
the spinons are deconfined and have a linear dispersion relation at small momenta. 
The spectrum is a gapless incommensurate multi-spinon continuum, described by TLSL 
theory at low energies. The incommensurability is directly related to the 
field-dependent magnetization \cite{Chitra1997}.

Experimentally, a field-induced deconfinement of spinons in the TLSL regime has been 
beautifully demonstrated in neutron scattering experiments and numerical simulations 
on the strong-rung ladder material (C$_5$H$_{12}$N)$_2$CuBr$_4$ (BPCB) 
\cite{Thielemann2009p2,Bouillot2011p1}.  In such strong-rung ladders, due to the 
large gap and small bandwidth, spectral components arising from the three members 
of the initial triplet of magnons remain energetically separated in a wide range 
of fields above $H_{\mathrm{c}1}$. In particular, the lower-energy excitations 
studied in BPCB \cite{Thielemann2009p2} are all descendants of the $M_S=+1$ magnon 
mode. The main idea of the present work is to study the field evolution of excitations 
in the strong-leg case. Here the spin gap is much smaller than the excitation bandwidth, 
and components of different spin polarization are no longer energetically separated. 
Since transitions between different spin states now occur at relatively low energies, 
the resulting spectrum is expected to be considerably more complex with numerous 
interesting high-energy features.

We study the compound (C$_7$H$_{10}$N)$_2$CuBr$_4$ (DIMPY) \cite{Shapiro2007p1,Hong2010p2}, 
which is structurally somewhat similar to BPCB. Unlike BPCB though, it realizes the 
{\it strong-leg} AF Heisenberg $S=1/2$ ladder model. As described in \cite{Hong2010p2,Schmidiger2011p1}, 
the ladders are built of Cu$^{2+}$ and run along the  $a$-axis of the monoclinic crystal structure. 
Previous experiments and their comparison to DMRG calculations established the spin Hamiltonian \cite{Schmidiger2011p1,Schmidiger2012p1,Schmidiger2013p1}. The bulk of all experimental data are 
adequately reproduced assuming just two Heisenberg exchange constants, $J_\mathrm{leg} = 1.42$~meV 
and $J_\mathrm{rung} = 0.82$~meV for the ladder legs and rungs. Such small energy scales, much 
smaller than those in the known strong-leg ladder compound La$_4$Sr$_{10}$Cu$_{24}$O$_{41}$ \cite{Notbohm2007}, 
make the quantum phase transition in DIMPY at $H_{\mathrm{c}1}\simeq 2.6$~T easily accessible in neutron experiments.

%
%
\begin{figure}[h!t]
\centering
\includegraphics[width=1\columnwidth]{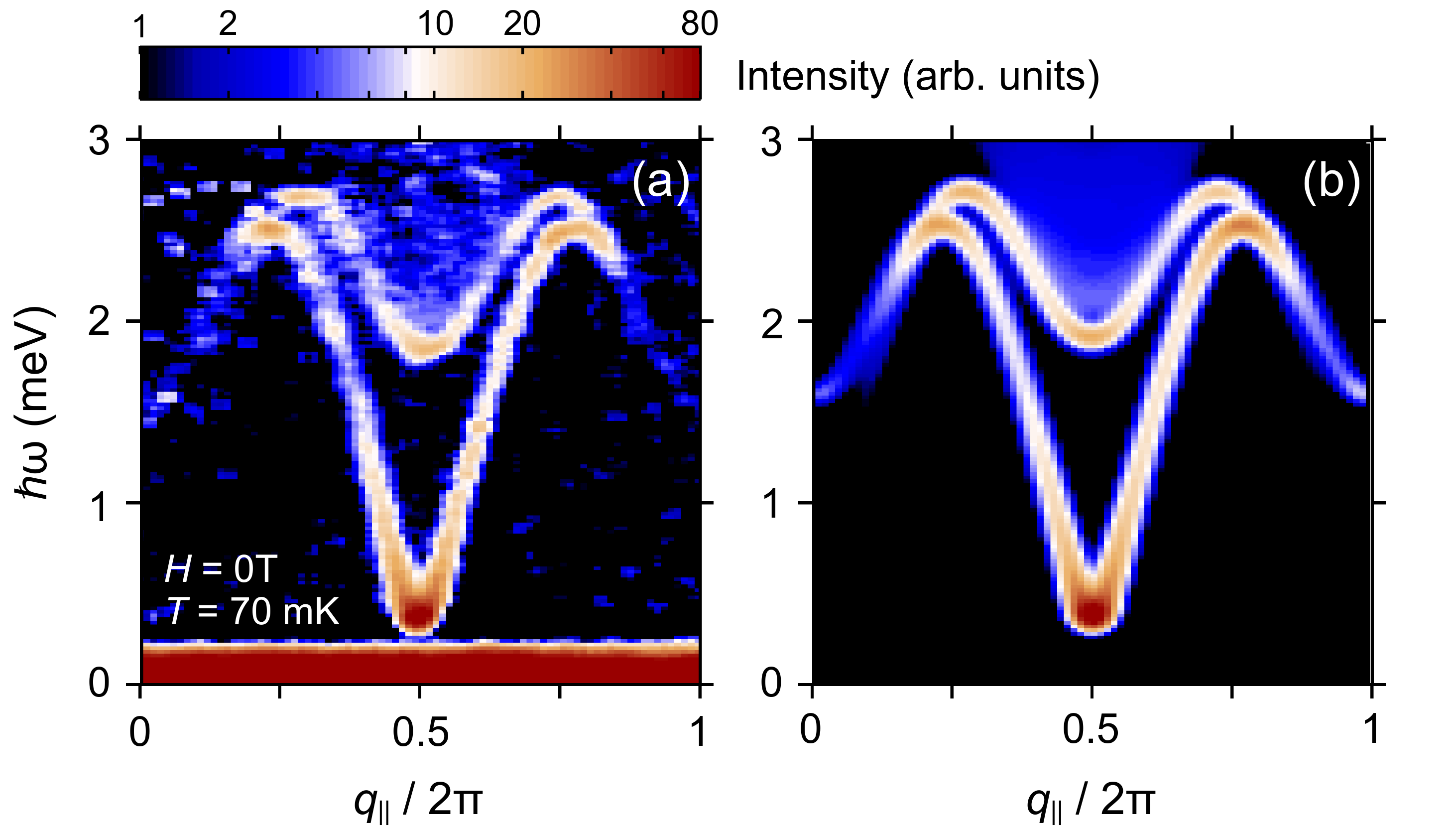}
\caption{(Color online) (a) Time of flight inelastic neutron scattering spectrum measured on 
DIMPY in zero applied field  ($E_\mathrm{i}=4.2$~meV). (b) A DMRG calculation of the same 
spectrum, folded with the known resolution function of the neutron instrument for a direct 
comparison with experiment.} \label{fig1}
\end{figure}

%

As a reference point, in Figure \ref{fig1}a we show the magnetic excitation spectrum measured 
in DIMPY in zero applied field at $T=70$~mK.  These data were taken on the same sample as in 
previous studies \cite{Schmidiger2011p1,Schmidiger2012p1,Schmidiger2013p1}, using the LET
neutron time-of-flight spectrometer at ISIS \cite{Bewley2011p1}. The sample was mounted on a 
$^{3}$He-$^{4}$He dilution refrigerator inside a vertical cryomagnet. Taking advantage of 
repetition rate multiplication, low- and high-resolution spectra with an initial energy of 
$E_\mathrm{i} = 4.2$~meV and $2.2$~meV were recorded at the same time. The inelastic 
background originating from the sample and instrument components was determined from 
the zero-field measurement and subtracted pixel by pixel \footnote{Having studied the 
zero-field spectrum in great detail, we assumed that spectral regions without expected 
magnetic signal correspond to non-magnetic background contributions. By linear interpolation, 
the complete background was calculated.}. All data shown are integrated along the 
non-dispersive $b^\star$ and $c^\star$ directions and projected onto the plane 
defined by energy transfer $\hbar \omega$ and momentum transfer along the leg 
direction $q_{||}=\mathbf{Q\!\cdot\!a}$ {\cite{Schmidiger2013p1}}.

As discussed in detail in \cite{Schmidiger2012p1,Schmidiger2013p1}, this spectrum shows several 
distinct features. The lowest-energy excitations are a triplet of single-magnon states that are 
the key prediction of the NLSM \cite{Shelton1996p1}. In DIMPY they have a gap of $\Delta=0.33$~meV 
at $q_{||}=\pi$ and a dispersion with a relativistic velocity $c=2.36$~meV \footnote{In this 
work we measure the velocity of spin excitations in energy units, as defined by a linear 
one-dimensional dispersion relation written as $\hbar \omega =c \mathbf{Q\!\cdot\!a}$. The 
values of $\Delta$ and $c$ were extracted from the previously measured dispersion relation 
in Ref. \onlinecite{Schmidiger2011p1}}. At higher energy transfers, there is an extended 
two-magnon bound state separated from a diffuse mutli-magnon continuum at even higher 
energies. While both a continuum and bound state \cite{Schmidiger2013p1} are predicted by 
the field-theoretical approach of \cite{Shelton1996p1}, it is not able to describe their 
dispersion and internal structure. The experiment is in almost perfect quantitative agreement 
with DMRG calculations (Fig. \ref{fig1}b) based on just the two exchange constants quoted 
above. Here the numerical result was convoluted with the instrumental resolution, allowing 
a direct comparison  to the experimental data.

\begin{figure}[h!t]
\centering
\includegraphics[width=\columnwidth]{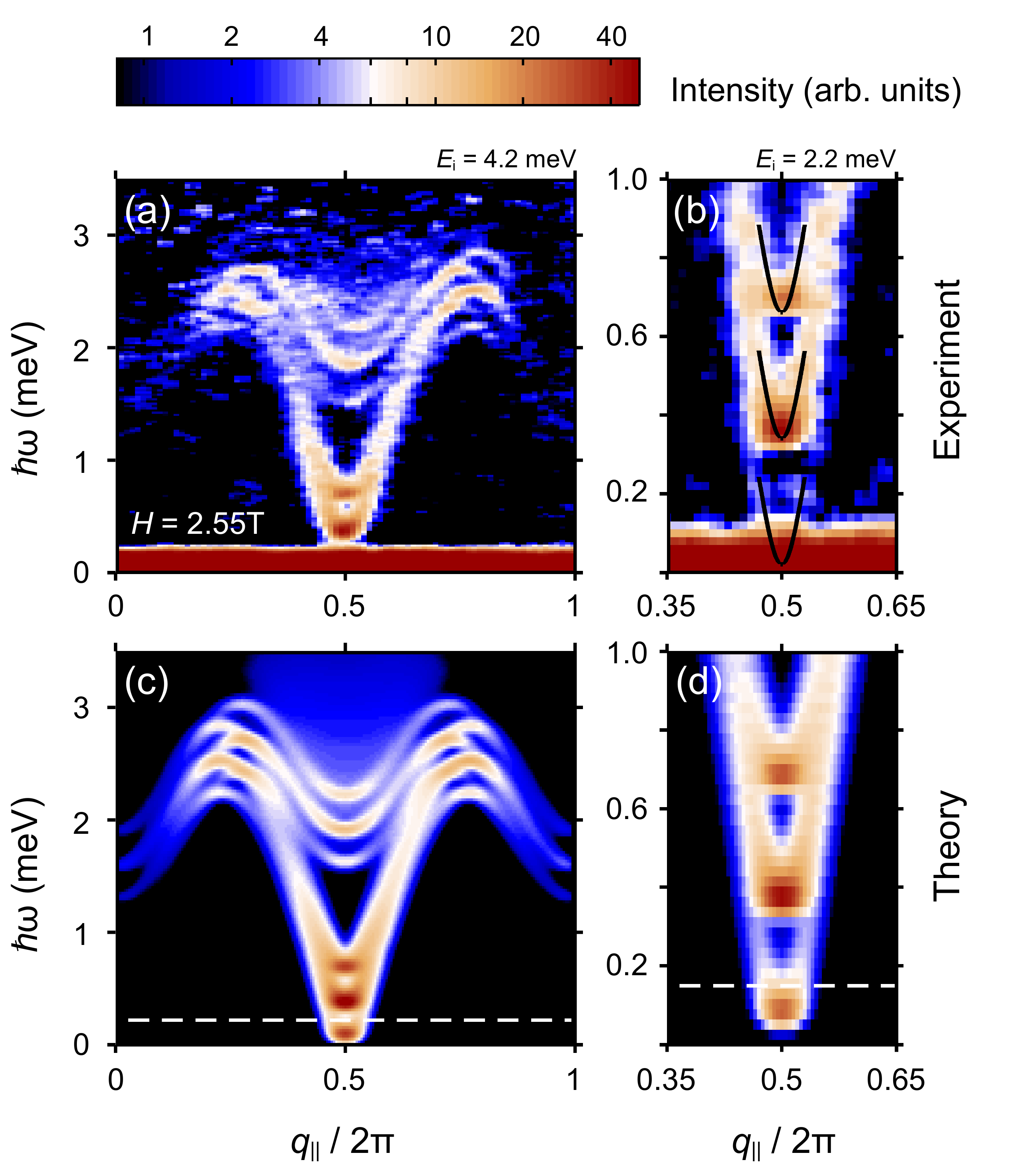}
\caption{(Color online) Spin excitations in DIMPY at $H = 2.55$~T. Inelastic neutron 
data were measured at $T=70$~mK in the low-resolution (a: $E_\mathrm{i} = 4.2$~meV) 
and high resolution (b: $E_\mathrm{i} = 2.2$~meV) modes. (c) and (d) are numerical 
DMRG calculations, convoluted with experimental resolution. Dashed lines indicate 
the onset of the elastic line in the experiment. Solid lines indicate the low-energy 
magnon excitations with relativistic dispersion.} \label{fig2}
\end{figure}

An external magnetic field $H$ lifts the threefold degeneracy of the
excitation spectrum. This is clearly visible in Figure~\ref{fig2}a,b, which  
summarizes the data collected at $H = 2.55$~T~$<H_{\mathrm{c}1}$ applied along 
the crystallographic $b$ axis in low- (a) and high-resolution (b) setups.
As long as the critical field is not exceeded, the singlet ground state of the 
spin ladder remains unchanged. Since the Zeeman term commutes with the Heisenberg 
Hamiltonian, the spectrum can be viewed as a superposition of three distinct 
polarization channels with excitations carrying $M_S = 0,\pm1$, respectively, 
$M_S$ being the eigenvalue of the $S_z^\mathrm{tot}$ operator and $z$ the direction 
of applied field. Since the temperature is low enough to ensure a negligible 
population of excited states, the three components are identical to the spectrum 
at $H=0$, except for an overall Zeeman energy shift. This applies not only to 
the magnon branch, but to all features, including the bound states and continua. 
In a neutron scattering experiment, the relative intensities of the three 
polarization contributions 
%
%
%
%
scale as $1:2:1$, for $M_S=-1$,$0$, and $1$, 
respectively, due to the intrinsic polarization dependence of the neutron 
scattering cross section. A DMRG calculation (fig. \ref{fig2}c,d) for the 
same conditions as the neutron experiment illustrates this simple yet important result.

\begin{figure}[h!t]
\centering
\includegraphics[width=1\columnwidth]{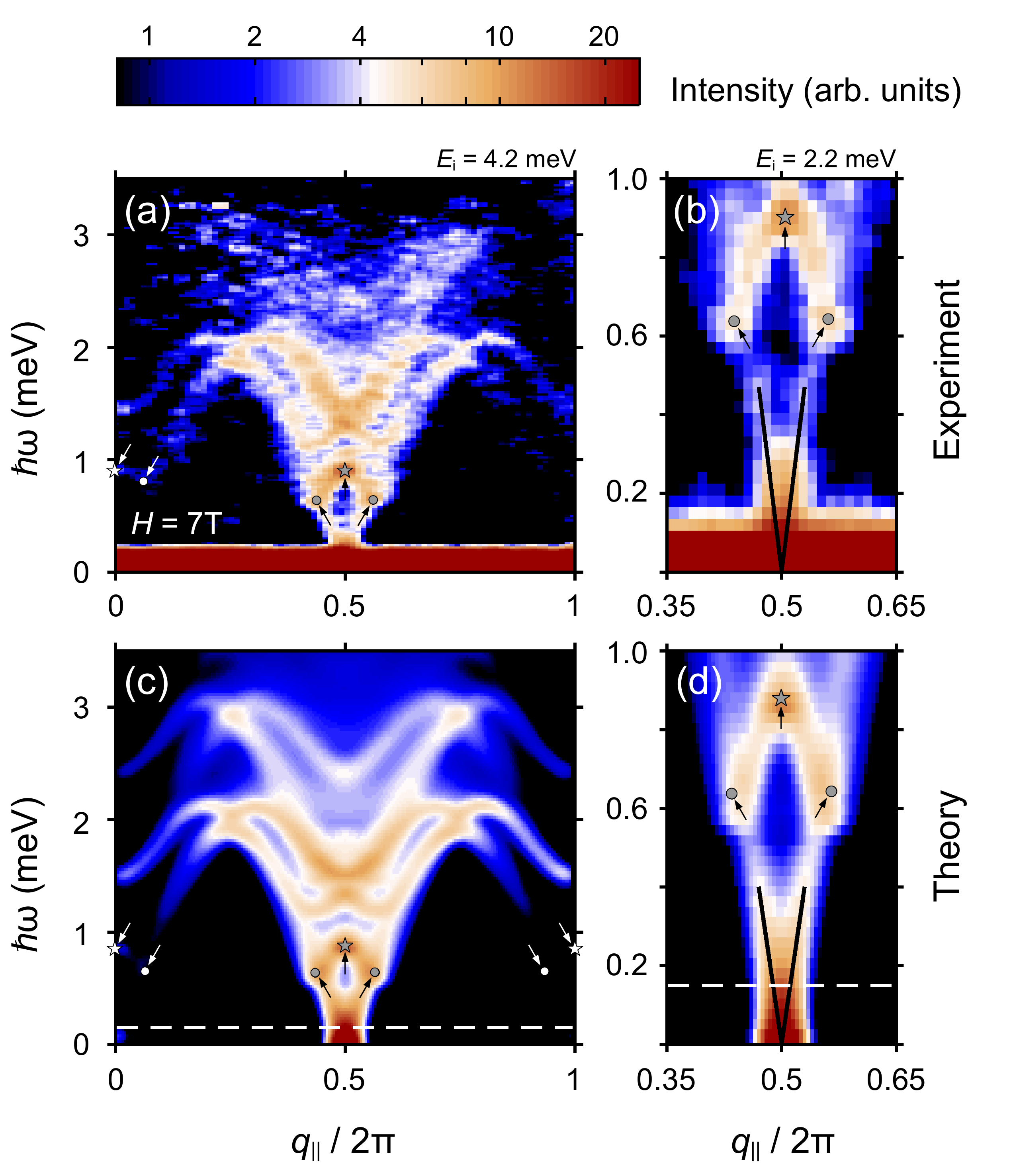}
\caption{(Color online) Spectrum of DIMPY at $H = 7$~T. Panels (a)--(d) and dashed 
lines are as in Fig.~\ref{fig2}. Lines are linear fits to the dispersion of the 
gapless excitations. Symbols { with arrows} are specific features predicted by a 
QFT mapping of the ladder model \cite{Furusaki1999p1}, as described in the text.}  
\label{fig3}
\end{figure}

The central result of this study is the measurement of excitations in the gapless 
TLSL phase. As can be seen from the data measured in DIMPY at $H=7$~T 
(Fig.~\ref{fig3}a,b), beyond the critical field the spectrum undergoes 
qualitative changes compared to that at low fields. The previously sharp modes 
decompose into structured and overlapping continua. A gapless linearly dispersive 
excitation emanating from $q_\| = \pi$ is seen at low energies. Many more distinct 
gapped features, with minima at either commensurate or incommensurate wave vectors, 
appear at higher energies. 

%
%
%

To make sense of this multitude of spectral features, we are going to classify 
them by their spin projections and by their parity with respect to a permutation 
of the equivalent ladder legs. Both quantities are conserved by the Heisenberg and 
Zeeman Hamiltonians. One has to keep in mind though, that at $H>H_{\mathrm{c}1}$ 
the spin ladder is magnetized, and therefore the contributions of different 
polarization are no longer identical. The total neutron scattering cross section 
can be broken up into six independent parts \cite{Bouillot2011p1}:
\begin{align}
\frac{\mathrm{d}^2\sigma}{\mathrm{d}\Omega \mathrm{d}E'} \propto  
4\biggl(1-\frac{Q_z^2}{Q^2}\biggr) [{s^{+}}({\bf Q})\, S_{0}^{zz}+{s^{-}}
({\bf Q})\, S_{\pi}^{zz}]+\notag\\ +\biggl(1+\frac{Q_z^2}{Q^2}\biggr)
[{s^{+}}({\bf Q}) (S_{0}^{+-}+S_{0}^{-+})+{s^{-}}({\bf Q})\,(S_{\pi}^{+-}+S_{\pi}^{-+})].
\label{eq}
\end{align}
Here, $Q_z$ denotes the component of the momentum transfer ${\bf Q}$ along magnetic field. $\mathcal{S}_{q_\perp}^{\alpha\beta}=\mathcal{S}_{q_\perp}^{\alpha\beta}(q_\|,\omega)$ 
are dynamic correlation functions associate{d} with the different symmetry channels.
The superscripts $\alpha$ and $\beta$ label the correlated spin components: 
$(\alpha,\beta) =(zz), (+-), (-+)$. The leg permutation parity $p=1,-1$ is 
represented by the momentum transfer along the ladder rungs $q_\perp $, defined 
as $p=\exp (\mathrm{i} q_\perp)$.  In this notation $q_\perp =0,\pi$ corresponds 
to symmetric and asymmetric excitations, respectively. ${s^{-}}({\bf Q})$ and 
${s^{+}}({\bf Q})$ are the asymmetric and symmetric structure factor, as 
discussed in detail in \cite{Schmidiger2011p1,Schmidiger2013p1}.

\begin{figure}[h!t]
\centering
\includegraphics[width=1\columnwidth]{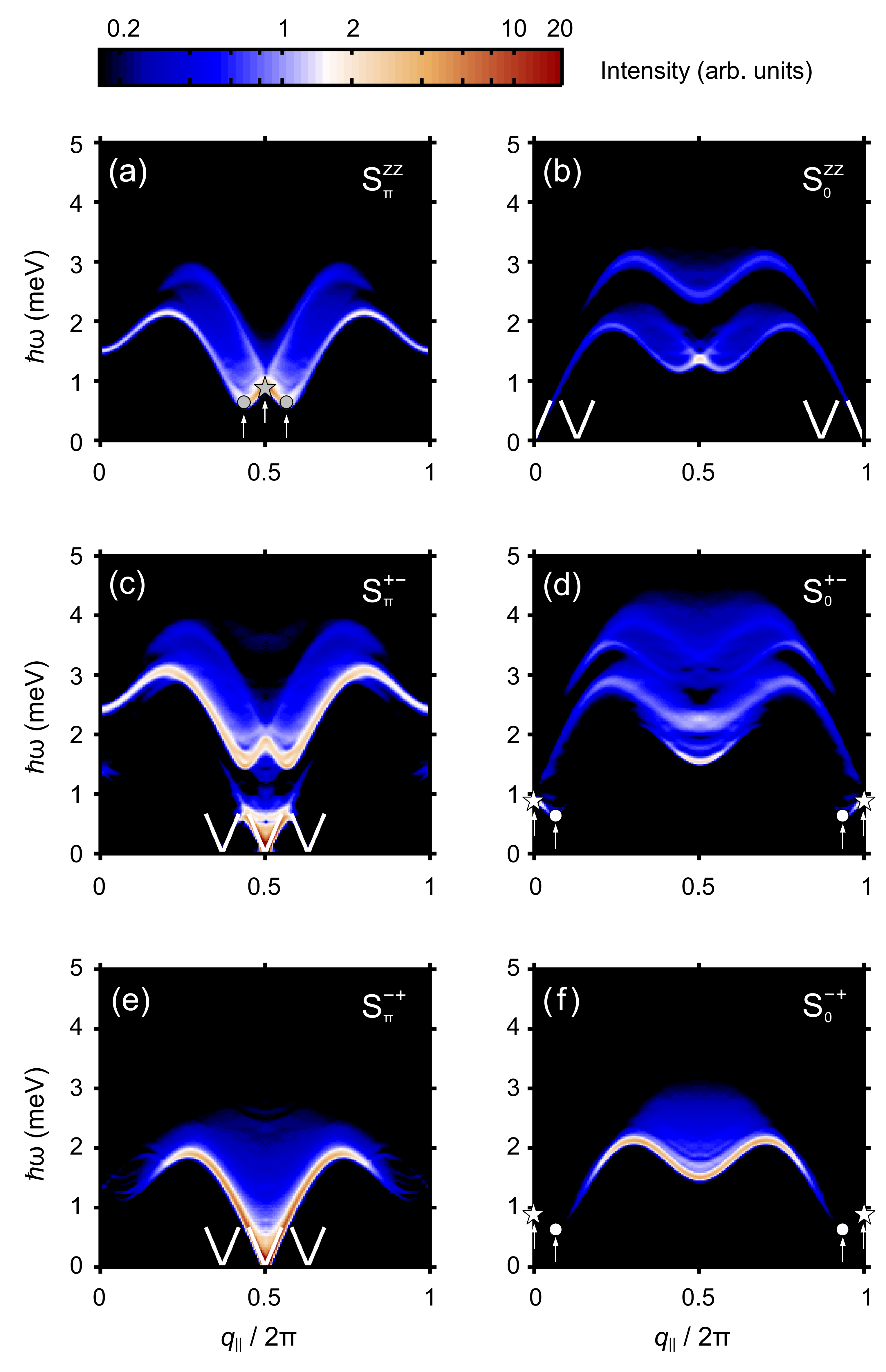}
\caption{(Color online) Calculated components of the $H=7$~T excitation 
spectrum of DIMPY, classified by their spin projection quantum number 
(top to bottom) and parity with respect to ladder leg interchange (left: 
asymmetric, right: symmetric). Lines and symbols are as in Fig.~\ref{fig3}.} 
\label{fig4}
\end{figure}

It is, in principle, possible to separate the six channels in inelastic 
neutron scattering experiments, by performing measurements at different 
wave vectors in different Brillouin zones or by applying a horizontal 
magnetic field. In practice, this procedure is extremely challenging. Instead, 
to identify the various spectral components in the experimental data, 
we took guidance from DMRG calculations. In such simulations, the 
individual contributions $\mathcal{S}_{q_\perp}^{\alpha\beta}$ can 
be accurately obtained. For DIMPY, $H=7$~T and $g=2.17$ \cite{Schmidiger2012p1} 
for Cu$^{2+}$, the result is shown in Fig.~\ref{fig4} \footnote{The DMRG 
computation parameters were the same as in Ref. \onlinecite{Schmidiger2012p1}}. 
It corresponds to a net magnetization per site of $\langle S_z \rangle = 0.065$. 
The complete calculated cross section, for a direct comparison with experiment, 
is shown in Fig.~\ref{fig3}c,d. The spectacular agreement with the neutron data 
gives us confidence that the measured spectra can be deciphered using the 
numerical ``key'' of Fig.~\ref{fig4}.

%
%
%

Our goal is, wherever possible, to relate the features observed to those predicted 
by the field-theoretical mapping of  Ref.~\cite{Chitra1997,Shelton1996p1,Furusaki1999p1}. 
Starting with low energies, at $H>H_{\mathrm{c}1}$, one expects several gapless excitations, 
generic to the TLSL state \cite{Chitra1997,Furusaki1999p1}. These continua are descendants of 
the soft {$M_S=+1$} magnon. They have a linear lower thresholds of 
$(\hbar \omega)^2 = v^2 (q_\|-q^\star)^2$, as indicated by solid lines in Figs.~\ref{fig3} 
and \ref{fig4}. Commensurate gapless excitations with $q^\star = \pi$ occur in the 
$\mathcal{S}^{\pm \mp}_{\pi}$ channel and are readily observed experimentally.
Incommensurate excitations around $q^\star = \pi \pm 4\pi \langle S_z\rangle$
and $q^\star = \pm 4\pi \langle S_z\rangle$ are predicted in the 
$\mathcal{S}^{\pm \mp}_{\pi}$ and $\mathcal{S}^{zz}_{0}$ channel respectively. 
They are detected by our DMRG calculations, but appear at least 2 orders of magnitude 
weaker than the commensurate ones, and therefore remain undetected in our experiments. 
Contrarily, gapless incommensurate excitations were observed in strong-rung 
ladders \cite{Thielemann2009p2}. This remarkable difference can be explained by the 
different nature of interactions between spinons (spinons are repulsive in strong-rung 
and attractive in strong-leg ladders), leading to distinct TLSL exponents $K$ but also
to different amplitudes of commensurate and incommensurate excitations \cite{Bouillot2011p1}. 
The ratios $(\alpha_1,\alpha_2) = (A_x/A_z,A_x/B_x)$ between commensurate (incommensurate) 
excitations in $\mathcal{S}^{\pm\mp}_{\pi}$ with amplitude $A_x$ $(B_x)$ and incommensurate
excitations in $\mathcal{S}^{zz}_{0}$ (amplitude $A_z$) at $\langle S_z \rangle = 0.065$ were 
numerically calculated to be $(\alpha_1,\alpha_2) = (13.1,8.9)$ for 
DIMPY \cite{Schmidiger2012p1}\footnote{The amplitudes $A_z$ and $B_x$ were calculated as in 
Ref. \cite{Schmidiger2012p1} but not shown there. { The value of the TLSL exponent $K$ 
at $\langle S_z \rangle = 0.065$ is $K=0.966$ and $1.170$ for BPCB and DIMPY respectively.}} and
$(2.5,1.3)$ for BPCB \cite{Bouillot2011p1}, proving that incommensurate contributions are 
suppressed in strong-leg ladders. For the commensurate modes in DIMPY, linear fits of the 
experimental and numerical low-energy spectra, allow us to extract the TLSL 
velocity $v$ \cite{Bouillot2011p1}. Experimentally, we obtain $v = 2.5(2)$~meV, which 
is comparable to $v = 2.0 (1)$~meV as estimated {\footnote{The fit errors are estimated 
using different fitting procedures.}} from the calculated spectrum 
$\mathcal{S}_{\pi}^{-+}$ or $v = 1.87(1)$~meV from static correlations in \cite{Schmidiger2012p1}.

In addition to the gapless spectrum, Ref.~{\cite{Chitra1997,Furusaki1999p1}}
predicts {\itshape gapped} excitations that are descendants of the $M_S = 0$ 
magnon. They appear in the $\mathcal{S}^{zz}_{\pi}$ as well as in the 
$\mathcal{S}^{\pm \mp}_{0}$ channel and show incommensurate minima at 
$q_\|=\pi \pm 2\pi \langle S_z\rangle$ (gray circles in Figs. \ref{fig3}, \ref{fig4}) 
and $q_\|= \pm 2\pi \langle S_z\rangle$ (white circles in Figs. \ref{fig3}, \ref{fig4}) 
respectively. The minima around $q_{||} = \pi$ are readily observed in our measured 
spectrum (grey circles in Fig.~\ref{fig3}). They are located at $q_\|/2\pi = 0.435(3)$ 
and $0.560(3)$, in excellent agreement with expectation based on the known value of 
$\langle S_z\rangle$. Moreover, we observe a hint of an incommensurate minimum at 
$q_\|/2\pi = 0.06(2)$ (white circle in Fig.~\ref{fig3}a) which agrees with
the predicted excitation at $q_\|= \pm 2\pi \langle S_z\rangle$ in the 
$\mathcal{S}^{\pm \mp}_{0}$ channel. However, while we reproduce the predicted 
$q_{||}$-position of the incommensurate gapped minima, we observe an additional 
energy shift due to a renormalization of the gap outside the vicinity of the 
quantum critical point.

Quantum field theory ~\cite{Furusaki1999p1} predicts that excitations at the 
zone center $q_{||} = 0$ (in $\mathcal{S}^{\pm \mp}_{0}$) as well as at 
the magnetic zone center $q_{||} = \pi$ (in $\mathcal{S}^{zz}_{\pi}$) appear 
directly at $\hbar \omega = g\mu_\mathrm{B} H$. In our data on DIMPY, we observe 
corresponding excitation energies of $\hbar \omega = 0.90(1)$~meV at $q_{||} = \pi$ 
(grey stars in Figs. \ref{fig3},\ref{fig4}) and $\hbar\omega = 0.896(4)$~meV 
at $q_{||} = 0$ (white stars in Figs. \ref{fig3},\ref{fig4}), in excellent 
agreement with $g\mu_\mathrm{B}H = 0.89$~meV for $g=2.17$ and $H = 7\,$T. 

We hence experimentally verify most of the QFT predictions 
in Refs.~{\onlinecite{Chitra1997,Furusaki1999p1}}. However, many of the observed 
features remain unaccounted for. These include the internal structure of the 
continua steming from the {$M_S = 0$} magnon branch, as well as the descendant of 
the $M_S = -1$ magnon in the $\mathcal{S}^{zz}_{\pi}$ and $\mathcal{S}^{+-}_{\pi}$ 
channel respectively (fig. \ref{fig4}a,c). Moreover, the existing QFT results do 
not bring any insight on how two-magnon excitations evolve in the TLSL regime. 
In Figures~\ref{fig4}b,d and f, remains of the two-magnon bound state and continuum 
excitations are still visible in the $\mathcal{S}^{-+}_{0}$, $\mathcal{S}^{zz}_{0}$ 
and $\mathcal{S}^{+-}_{0}$ channels. Apart from a shift in energy, a slightly reduced 
bandwidth and a loss of intensity, two-magnon excitations in the $\mathcal{S}^{zz}_{0}$ 
and $\mathcal{S}^{-+}_{0}$ channel resemble the corresponding contribution in zero field. 
At the same time, two-magnon excitations in the $\mathcal{S}^{+-}_{0}$ channel 
(at highest energies in fig. \ref{fig4}d) gain a novel internal structure.

In conclusion, we were able to measure and numerically calculate the complete spin
excitation spectrum of a strong-leg spin ladder material in both the gapped and 
TLSL phases. Predictions of continuous quantum field theories are verified on the 
quantitative level. At the same time, many of the observed higher-energy spectral 
features are not captured by that powerful, yet intrinsically limited, 
low-energy approach.

This work is partially supported by the
Swiss National Fund under division II and
through Project 6 of MANEP. DS and AZ would 
like to thank S. M\"uhlbauer (FRM II, Technische Universit\"at
M\"unchen) for his involvement at the early stages of this 
project. Finally, we thank the sample environment team of the ISIS 
facility for their excellent support during the experiment.



\end{document}